# Modular molecular toolkit for photochemical energy conversion in a self-assembling nanocontainer

*David Kaftan[1,2], David Bína[1,3], Guy Michel Wolf[1], Martin Baroch[4,5], Jakub Ködel[1], Juraj Dian[4,5], Filip Dyčka[1], J. Thomas Beatty[6], Heiko Lokstein[5], Jakub Pšenčík[5*] and Roman Tuma[1*]*

[1] University of South Bohemia, Department of Chemistry, Branišovská 1760, 37005 České Budějovice, Czech Republic

[2] Czech Academy of Sciences, Institute of Microbiology, Centre Algatech, Novohradská 237, 37981 Třeboň, Czech Republic

[3] Czech Academy of Sciences, Biology Centre, Institute of Plant Molecular Biology, Branišovská 31, 37005 České Budějovice, Czech Republic

[4] Charles University, Faculty of Science, Department of Analytical Chemistry, Albertov 6, 12843 Prague, Czech Republic

[5] Charles University, Faculty of Mathematics and Physics, Department of Chemical Physics and Optics, Ke Karlovu 3, 12116 Prague, Czech Republic

[6] The University of British Columbia, Department of Microbiology & Immunology, Vancouver, V6T 1Z3, Canada

KEYWORDS: capsid, *Cereibacter*, electron transfer, photosynthesis, P22 bacteriophage, nanobiotechnology

ABSTRACT

Production of useful chemicals using photoelectrochemical biohybrid devices offers an environmentally friendly alternative to existing energetically demanding processes. These devices exploit light-driven charge separation, e.g. by a photosystem, and require efficient electron transfer to a tailored redox enzyme cascade. Here we demonstrate that electron transfer efficiency can be increased by confining the photosystem with the redox protein inside a self-assembling, virus-based nanocontainer. The photosynthetic system from the phototrophic bacterium *Cereibacter sphaeroides* and cytochrome *c* were conjugated to a bacteriophage P22 scaffolding protein and co-incorporated into the 50 nm diameter virus shell *in vitro*. The porous shell confined the macromolecular components for efficient electron transfer while allowing free exchange of small electron mediators. Sustainable and accelerated light-driven electron transfer between the encapsulated components was confirmed by optical spectroscopy. This self-assembly system presents a versatile platform for developing nanoreactors that combine photosystems with complex redox pathways.

## INTRODUCTION

The development of renewable technologies for harnessing solar energy for fuel and high-value product (HVP) synthesis from $CO_2$ is a major technological challenge. Natural photosynthesis, the only sustainable global $CO_2$ removal process, captures solar energy by light-harvesting complexes (LHCs) and converts it into chemical energy in reaction centres (RCs) [1]. The primary processes of photosynthesis occur with high quantum efficiency, providing models for solar energy utilization [2, 3 3-6]. The photosynthetic complexes of RCs and LHCs (collectively called photosystems, PS) have been integrated into hybrid systems for HVP synthesis, electricity generation [2, 7-10] and sustainable production of hydrogen [11-15].

While the primary photochemical steps are efficient, subsequent reactions, such as $CO_2$ reduction to carbohydrates or alcohols, limit the overall efficiency [2]. This is partly because the subsequent redox reactions rely on electron transport which requires close proximity between the electron donor and acceptor. Photosynthetic systems evolved specific docking interfaces between the redox complexes and mobile electron carriers to achieve efficient electron transport. This often poses a problem when designing biomimetic systems encompassing components from heterologous organisms in new pathways, because of incompatible protein-protein interfaces that are difficult to re-design.

To deal with this problem, applications of photosynthetic complexes in photovoltaics or HVP production employ electron transfer mediators (e.g. ascorbate, quinones, cytochromes) immobilized on the electrode to facilitate electron transfer [16, 17]. Cytochromes *c*, which serve as the natural electron donors to many RCs, are routinely used in biohybrid devices, either co-immobilized with PS on electrodes [8, 10, 16, 17] or as soluble electron carriers [8, 18-20]. HVP production additionally requires immobilization of the redox enzymes on electrodes via covalent bonding or nano-structuring, including layering [3, 4]. However, this approach becomes difficult for engineering of more complex redox cascades that may involve electron transfer as well as substrate and product channeling.

One way to deal with this problem is to exploit hierarchical self-assembly and nanoscale integration, which would lead to tight coupling between RC complexes and redox enzymes due to spatial confinement. Viral capsids offer inner cargo space (15–50 nm) for encapsulating and confining nanoparticles and enzymes [21-25]. The *Salmonella* phage P22 procapsid system incorporates protein cargo via co-expression of bacteriophage P22 coat protein (CP) and engineered scaffolding protein (SP) in *E. coli* [22, 23]. Alternatively, cargo can be incorporated into the procapsid *in vitro* [26] by self-assembly of stoichiometric CP and SP, with truncated SP variants leaving enough inner space for cargo (e.g. quantum dots) [24], which is linked to the N-terminus of the 144 C-terminal amino acids of a truncated SP[27].

Here we aimed to demonstrate that nanoscale spatial confinement within a virus nanoparticle facilitates light-driven electron transfer between RC and a heterologous redox protein, bovine cytochrome *c* (Cyt *c*). The genetically engineered bacterial PS from *Cereibacter* (*C.*) *sphaeroides* (RC-LH1) was selected as a model system.[28-30] It consists of a core antenna and RC which has a truncated H-subunit and a hexahistidine tag (6xHis) at its C-terminus and retains electron transfer capabilities [31, 32]. A versatile metal affinity conjugation scheme for linking the RC-LH1 and other redox components with the phage scaffolding protein was designed, and the phage P22 self-assembly reaction was optimized for *in vitro* incorporation of detergent-solubilized, SP-conjugated RC-LH1 and the Cyt *c*. This system yielded capsids with stoichiometric amounts of engineered RC-LH1 and cytochrome in the interior. Time-resolved spectroscopy confirmed electron transfer from the Cyt *c* to the photo-oxidized RC within the capsid. The modular self-assembly toolkit enables nanoscale confinement of heterologous cargo, including solubilized membrane complexes, and opens the way for designing nanoreactors harbouring complex light-driven redox pathways.

## RESULTS

### ***Toolkit component engineering***

Because of poor yield of direct gene fusion between the SP and RC H protein, linkage was achieved using bi-functional PEG linkers and $Ni^{2+}$ mediated coordination between chemically introduced nitrilotriacetic acid (NTA) moieties and 6×His tags.

The Assembler-derived SP possesses an N-terminal 6×His tag [23] (Fig. 1a) which was used for $Ni^{2+}$ mediated coordination with the NTA moiety that was conjugated to Cyt *c* via chemical derivatization of surface exposed lysines with NTA-succinimidyl (NHS) ester PEG linker. The primary amine groups of the two surface-exposed lysine residues on Cyt $c_2$ reacted with the NHS, resulting in approximately half of the Cyt $c_2$ molecules being conjugated to the linker (designated C-SP, Fig. 1a) as shown by the increased mass (Supplementary Fig. S1).

A "reverse" coupling was employed for conjugating the SP with the detergent-solubilized RC-LH1 complex, which is 6xHis tagged at the H-subunit C-terminus. The N-terminal 6xHis was removed from SP by Super TEV protease (Fig. 1a and Supplementary Fig. 2). A single engineered cysteine residue at the N-terminus of SP was then reacted with the maleimide-PEG-linked NTA reagent (Fig. 1a and Supplementary Fig. 2) and conjugated to the His-tagged H-subunit of the RC-LH1 complex in detergent micelles.

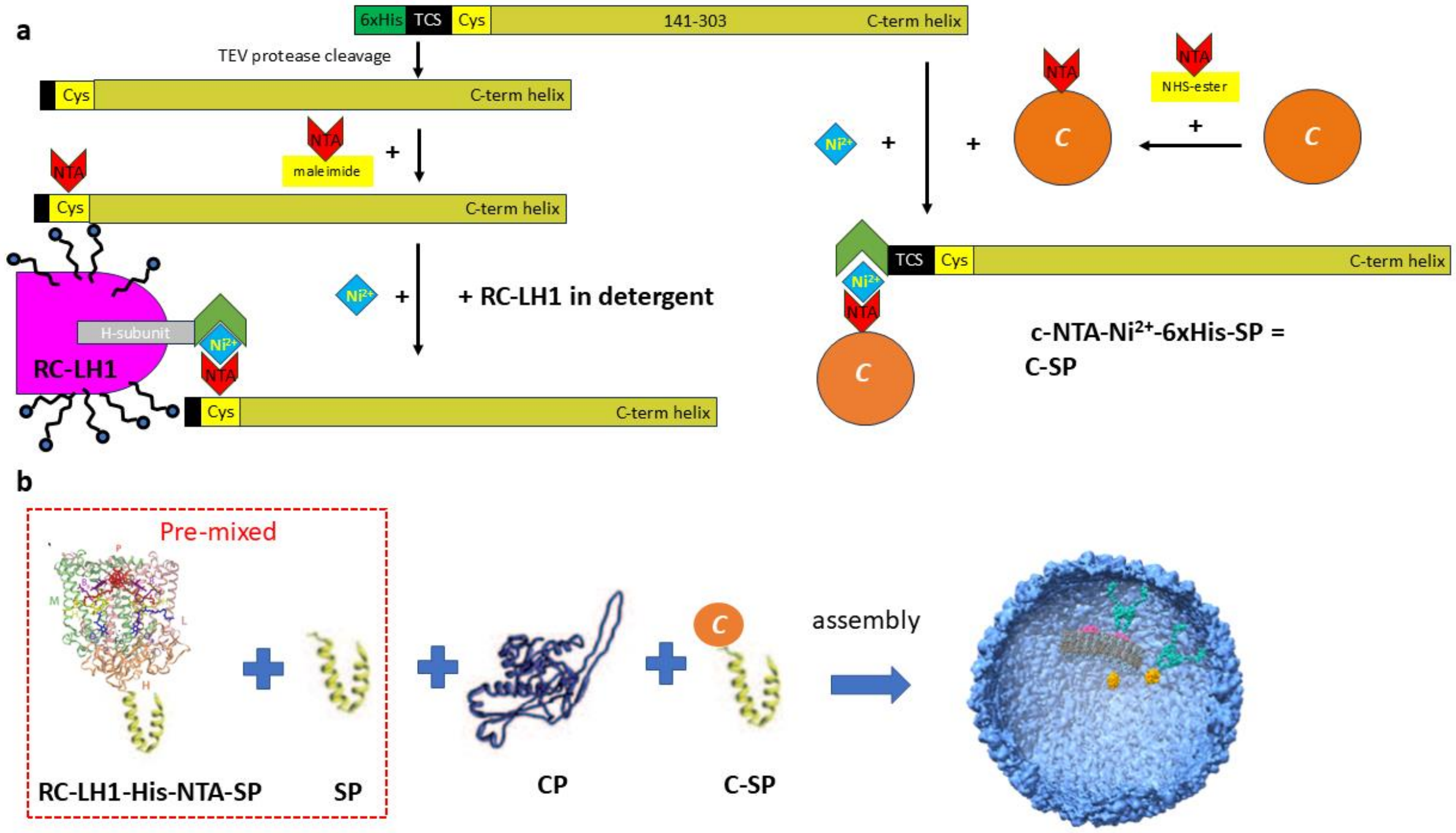

**Figure 1.** Schematics of assembly toolkit. (a) Scaffolding protein conjugation scheme. The minimal scaffolding fragment encompassing residues 141-303 (olive) was further engineered to contain a 6xHis tag at the very N-terminus (green) followed by TEV protease cleavage site (TCS - black) and single Cysteine (Cys – yellow). This precursor was cleaved with TEV protease and covalently linked with PEG-NTA via cysteine maleimide reaction (red/yellow - left). The NTA-conjugated SP was then complexed with RC-LH1 (pink) via 6xHis tag at the C-terminus (green) of truncated H-subunit (grey). Alternatively, the His-tagged version was complexed via Nickel metal affinity tag to an NTA-modified bovine Cyt *c* (orange - right). (b) Schematic representation of assembly. From left to right: RC-LH1-SP represented by RC structure (PDB 7VOY)[33] linked to the C-terminal SP helical domain (PDB 2GP8)[34] was premixed with excess of unmodified SP to prevent further oligomerization of RC-LH1 (red box) after which Cyt c and CP were added. An EM derived structure (PDB 2XYY)[35] was used to produce an artistic representation of the assembled procapsid shell (blue) with RC-LH1 (PDB 7VOT)[33] containing

micelle (grey) linked via SP (turquoise) to the inner surface together with Cyt *c* (orange). Note that the cartoons are not to scale.

***Nanocontainer assembly***

Purified procapsids, produced in *E. coli* from the P22-Assembler plasmid containing a TEV protease site inserted downstream from the N-terminal 6xHis encoded within the SP locus, served as a source of the capsid components for the *in vitro* assembly. The monomeric CP was obtained by denaturation and refolding of empty procapsid shells at concentrations up to 1 mg/ml that are sufficient for self-assembly and low enough to avoid losses to aggregation [36].

The incorporation of RC-LH1-SP into the procapsids was optimized by mixing the assembly reaction in dodecyl-β-D-maltoside (DDM) containing buffer, keeping the SP to CP molar ratio constant (SP:CP 1:2) and increasing the RC-LH1-SP conjugate to CP molar ratio in the reaction (from 1:400 to 1:20, Fig. 1b). The products were separated by HPLC and detected by absorbance at 280 nm (protein) and 500 nm (pigments, Fig. 2a). The first peak in the chromatogram corresponded to particles with MALS-derived $R_G$ radius of 22 ± 3 nm and DLS-derived $R_h$ 25 nm; i.e. those expected for approximately spherical P22 procapsid shells.

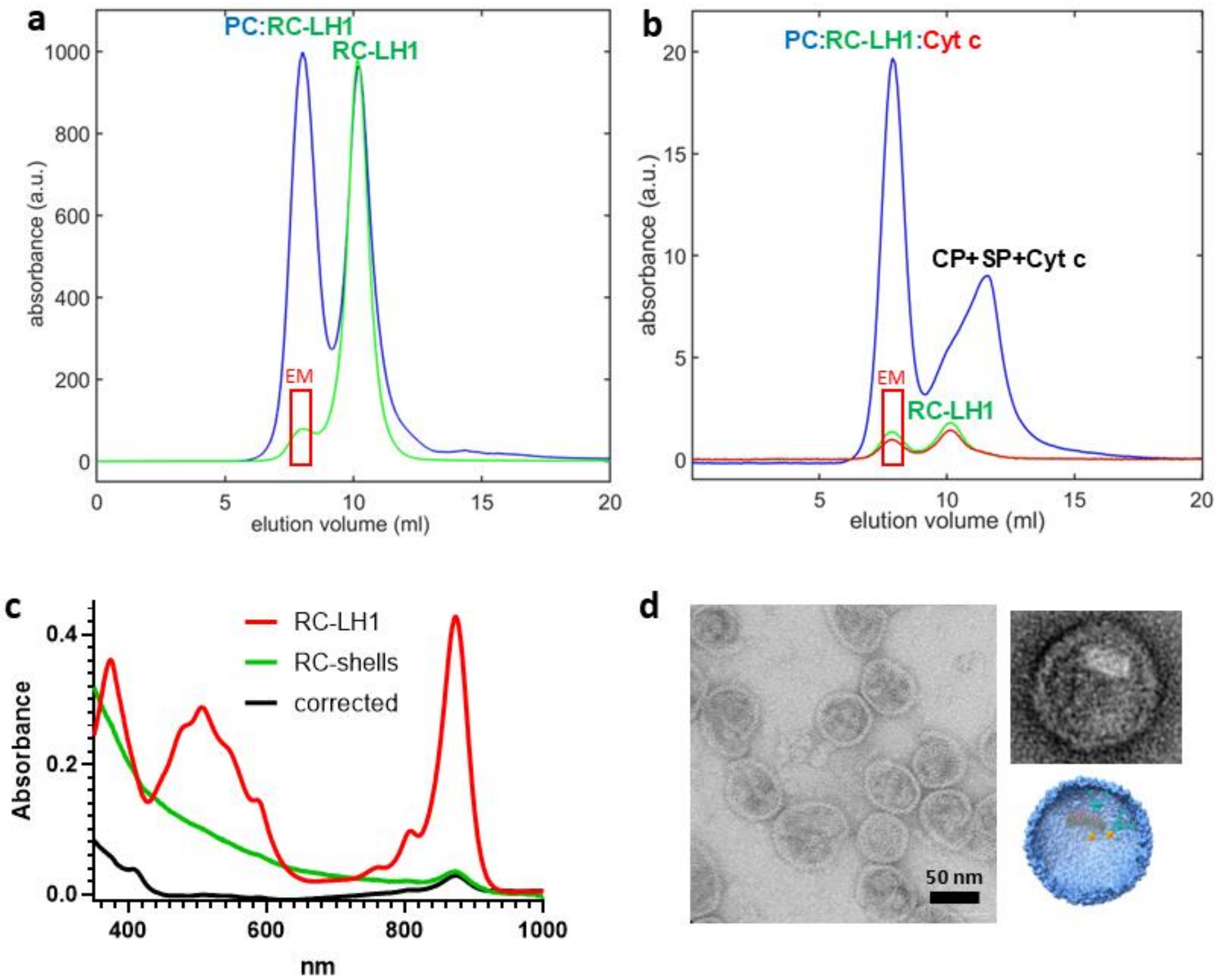

**Figure 2**. Assembly product characterization. (a-b) Chromatograms of assembly products monitored by absorbance at 280 nm (blue, proteins), 505 nm (green, RC-LH1 chlorophylls) and 550 nm (red, cytochrome c). Panel a – assembly of RC-LH1-SP; Panel b – co-assembly of RC-LH1-SP with Cyt c SP conjugate; PC – procapsid assemblies; Red box indicates peak fractions which were collected and examined by negative stain EM. Additional HPLC analyses of PC assemblies and components are shown in Supplementary Figure S3. (c) Optical absorption spectra of unassembled RC-LH1 (red) compared with HPLC purified capsids co-assembled with RC-LH1-SP (RC-shells, green) and scattering-corrected spectrum (black). (d) Negative stain EM field of view of assembly products from the PC:RC-LH1 peak fraction (left) and a representative detail containing RC-LH1 (upper right) which is compared with an EM-based model shown

below. EM micrograph gallery for assemblies from panels a and b is shown in Supplementary Fig. S4.

Both molar ratios and the sequence of addition proved important for incorporation and closed shell formation. Given the large size of RC-LH1, we aimed at incorporating up to 3 RC-LH1-SP per capsid to leave enough room for the cytochrome. Optimal incorporation resulted from mixing first 0.2 μM RC-LH1-SP with concentrated SP (~300 μM), then adding excess volume of CP monomer at 15-20 μM (to achieve 1:2 SP:CP molar ratio). Under those conditions about half of the coat protein assembled into shells (as judged by $A_{280}$ between the first and second peak, Supplementary Fig. S3a, red trace) while only a small fraction of the RC-LH1-SP was incorporated into the assembled products ($A_{500}$ nm, Fig. 2a, green trace).

Chromatography fractions containing the procapsid-size material yielded optical absorption spectra which exhibited the characteristic RC-LH1 band between 850-900 nm (Fig. 2c). Negative stain EM images (Fig. 2d) resembled the shape and size (50 nm diameter) expected for the procapsid (Fig. 2d, right panel and Supplementary Fig. S4a), some assemblies enclosed rectangular objects with dimensions consistent with the cryoEM structure of RC-LH1. However, as the fields of view show (Supplementary Fig. S4a), some of the particles were not fully closed or were malformed, suggesting heterogeneity. Quantitative proteomics for the presence of the desired proteins and their ratios (Supplementary Fig. S5) showed molar ratios of H/L/M RC subunits to CP ranging from $10^{-3}$ to $10^{-2}$: i.e., only one or few RCs per capsid on average.

The conditions optimized for RC-LH1 incorporation were used for co-encapsulation of RC-LH1 and Cyt *c*. NTA-functionalized Cyt *c* (45%, and 15% of Cyt *c* conjugated with 1, and 2 NTA linkers, respectively) was charged with $NiCl_2$ and mixed with SP in 1:9 ratio resulting in 100 μM Cyt *c*-SP conjugate, which was then mixed with SP in 1:19 molar ratio and added to a 10 μM CP solution. The final molar ratio of Cyt *c*-SP:SP:CP was 1:19:40. In order to achieve co-encapsulation of the RC-LH1 complex with the Cyt *c*, the SP coupled with the respective cargo were mixed in a molar ratio 1:2 (RC-LH1-SP:Cyt *c*-SP), and subsequently mixed with SP and monomeric CP in a final molar ratio 1:2:17:40 (RC-LH1-SP:Cyt *c*-SP:SP:CP). The assembly of the P22 procapsid-based nanocontainer loaded with cargo proceeded at 4 °C for 4 hours to prevent rapid formation of structurally aberrant procapsid shells. The assembled nanocontainers were purified using HPLC (Fig 2b) and the procapsid peak yielded absorbances at 500 and 550 nm compatible with incorporation of both RC-LH1 and Cyt *c* (reduced form). The peak fraction (Fig. 2b, red box) was subjected to negative stain EM which demonstrated presence of assembled particles with capsid morphology and heterogeneity similar to the RC-LH1-containing assemblies (Supplementary Fig. S4). The peak fractions corresponding to capsids containing RC-LH1 (Fig. 2a, red box) and the RC-LH1 with Cyt *c* (Fig. 2b, red box) were subsequently used for time-resolved spectroscopic measurements.

### *Facilitated photo-induced electron transfer within the nanocontainer*

The functionality of the primary photochemical events in the solubilized and encapsulated RC-LH1 complex was tested using kinetic absorption measurements (flash photolysis). The negative infrared signal between ~800-900 nm with a minimum at ~870 nm (bleaching of the RC primary donor $P_{870}$) and another minimum at 815 nm accompanied with a positive peak at ~790 nm (concomitant electrochromic blue shift of the accessory bacteriochlorophyll *a* (BChl) are characteristic of the oxidized RC [37]. The solubilized and encapsulated RC-LH1 complexes exhibited virtually indistinguishable flash photolysis absorption spectra, demonstrating the encapsulation had no effect on the primary photochemical events in the RC. The relaxation of the $P_{870}^{+}$ absorption change was then followed in RC-LH1 complexes in solution and encapsulated in the procapsids that were exposed to a single-turnover saturating flash (Fig. 3d). Again, no difference in the recovery kinetics was observed between solubilized and encapsulated RC-LH1 complexes. The addition of the $e^-$ donors (1 mM Na ascorbate, 3 μM PMS) increased the recovery to an equivalent rate in both the solubilized and encapsulated RC-LH complexes.

Finally, the relaxation of the $P_{870}^{+}$ absorption change was followed in the RC-LH1 encapsulated alone or co-encapsulated with Cyt *c* (in the absence of PMS) that were exposed to a 3 s long light pulse (Fig. 3d). The absorption changes of the encapsulated RC-LH1 fully recovered within tens of seconds after the end of the light pulse (Fig. 3d, red). The procapsids containing both the RC-LH1 and Cyt *c* showed a ~3-fold rate increase of $P_{870}^{+}$ reduction, demonstrating an enhancement of functional $e^-$ transport facilitated by the closer average distance and/or more frequent encounters between Cyt *c* and RC-LH1 due to the spatial confinement in the procapsid.

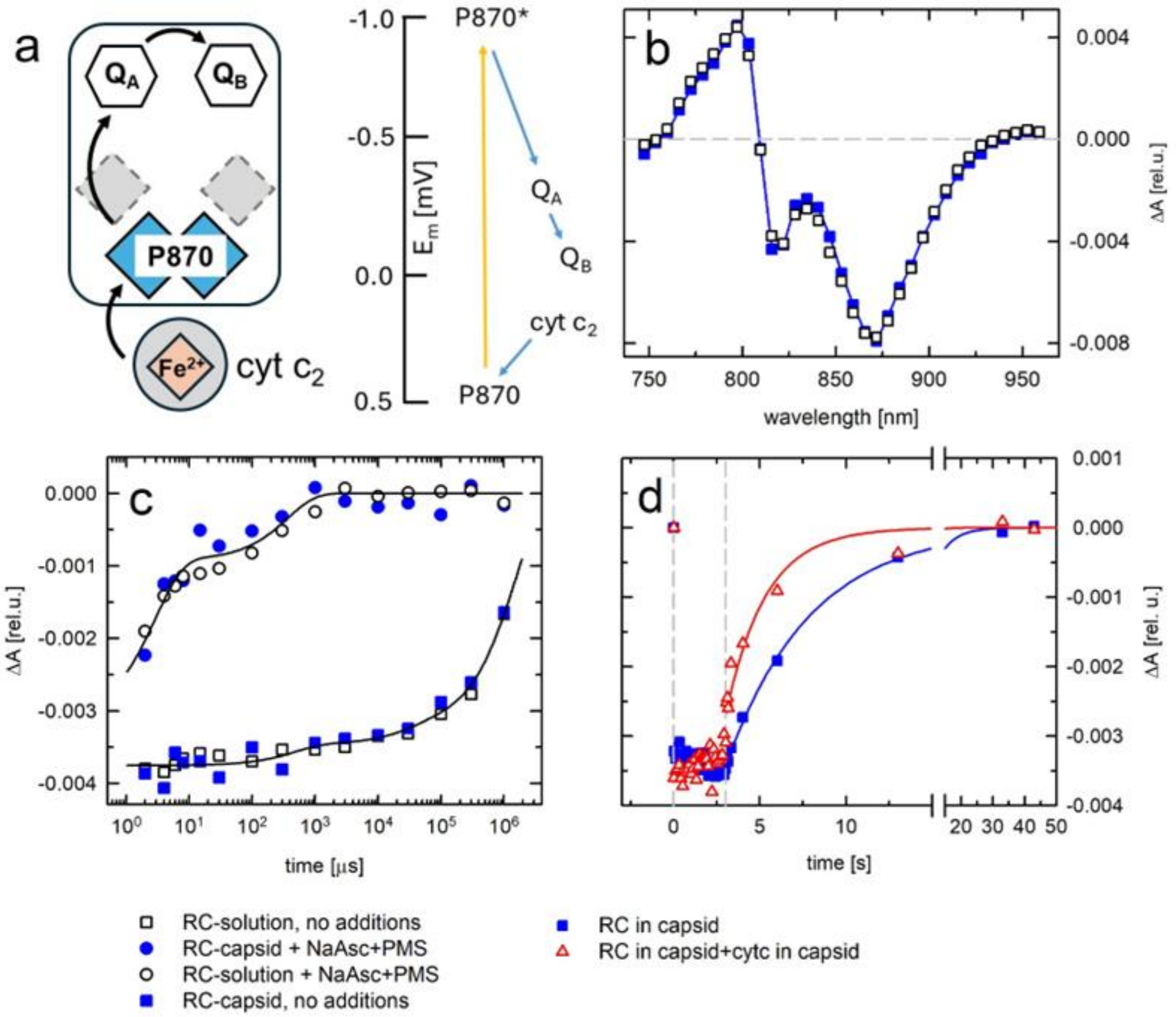

**Figure 3**. Electron transfer monitored by transient absorption spectroscopy. (a) A simplified scheme of the electron transfer processes and components and in the RC. P870 represents the primary donor, and $Q_A$ and $Q_B$ represent quinone electron acceptors. (b) Light minus dark differential absorption spectrum of RC-LH1 in solution (white squares) and enclosed in the P22 procapsid based nanocontainer (blue squares) demonstrates their unchanged photochemical conversion capacity. (c) Recovery of $P870^+$ after a 1 μs flash is accelerated by 4 mM Na-ascorbate and 3 μM phenazine methosulfate (circles) in RC-LH1 in solution (white) and in P22 procapsid based nanocontainer (blue); no e- donors (squares). (d) $P870^+$ recovery following a 3 s

actinic light pulse is accelerated in presence of Cyt c co-encapsulated in the P22 procapsid based nanocontainer together with RC-LH1 (red triangles); RC-LH1 encapsulated alone (blue); 4 mM Na-ascorbate present in all reactions. Additional data demonstrating reproducibility of measurements on different samples are shown in Supplementary Fig. S6.

## DISCUSSION

The standard architecture of biohybrid devices for solar energy conversion is based on co-immobilizing or layering PS and redox components onto electrodes [3, 4]. Amongst the limitations is inefficient electron transport between components resulting from sub-optimal interface engineering coupled with PS and redox components lacking spatial proximity. Here we show that spatial proximity can be achieved by confining a PS (the bacterial RC-LH1) and a redox component (bovine Cyt *c*) into the small internal volume of the bacteriophage P22 procapsid shell. The encapsulated components remain soluble without the need for detergent and resist dissociation. Using transient absorption spectroscopy we demonstrate that this approach yields a functional RC, and the capsid is permeable to small electron carriers (ascorbate and phenazine methosulfate). Thus, this approach is compatible with mediated electron transfer commonly used with electrodes (Cai et al. 2025). We have observed an enhanced electron transfer with co-encapsulated Cyt *c,* demonstrating accelerated intra-capsid electron transfer by spatial confinement. Previously, a rate enhancement of redox reactions upon packaging enzymes in the P22 capsid shell was observed, resulting from small electron carrier ($NAD^+$/NADH) channelling due to spatial confinement [38]. Here, we extended this result to demonstrate direct electron transfer between a PS, the large, detergent-solubilized bacterial membrane complex RC-LH1, and a heterologous bovine electron carrier Cyt *c*. Thus, spatial confinement is likely to enhance rates of both mediated and direct electron transfer, and also to yield enhancements due to channelling of small intermediates between enzymes, e.g. acids and aldehydes on the pathway to alcohols [3, 4].

Our approach to spatial confinement, which is based on modular component assembly (Fig. 1) and in vitro assembly, does not require direct genetic fusion of components. It allows for incorporation of membrane protein complexes, thereby expanding and complementing the P22 capsid assembler system, which is only suitable for soluble protein encapsulation by co-expression of genes in bacteria [39]. However, the in vitro assembly system also exhibits substantial drawbacks such as low efficiency of RC-LH1 incorporation, requiring higher than a stoichiometric ratio and thus limiting control over the stoichiometry of components in the final product.

## CONCLUSIONS

In summary the modular self-assembly system proposed here paves way for pre-assembling and spatially confining complex redox pathways including membrane proteins in a virus-based nanocontainers. Such nanoreactors can then be integrated into existing, electrode-based systems[3, 4, 40, 41,] for large scale solar production of chemicals from carbon dioxide in the future.

## MATERIALS AND METHODS

### ***Strains and growth***

*Escherichia* (*E.*) *coli* cells strain BL21 DE3* containing P22 cargo assembler plasmid coding for self-assembling *Salmonella thyphimurium* P22 virus procapsid was a kind gift from Trevor Douglas (O'Neil et al., 2011; Addgene plasmid # 122651; http://n2t.net/addgene:122651; RID:Addgene_122651). The plasmid sequence was additionally modified by PCR mutagenesis using primer1_fwd: TGTACTTTCAGAGCtgtcgcagcaatgcc, and primer1_rev: GATTTTCGCCGCTgctgtggtgatgatgg, replacing the thrombin cleavage site downstream of the N-terminal 6× His with the TEV (Tobacco Etch Virus) protease cleavage site. Cells were grown in 0.8 L baffled bottom Erlenmeyer flasks in LB medium at 37 °C, aerated in an orbital shaker at 180 rpm. Cultures that reached $OD_{500}$=1 were chilled on ice for 1 hour, followed by an induction of the expression of the procapsid proteins by 1 mM Isopropyl β-d-1-thiogalactopyranoside (IPTG). Protein production then proceeded with resumption of shaking at 37 °C for 4 hours.

Electrocompetent cells of *C. sphaeroides* strain $RC_x^r$ [30] were prepared according to Donohue & Kaplan [42]. Briefly, exponentially growing cells were centrifuged three times at 4 °C for 10 minutes at 4,000×*g*, and resuspended into ice-cold sterile deionized water, and finally electroporated by a single pulse of 8.5 ms in a 2 mm electrode gap cuvette (400 ohm) at 2.5 kV, 25μF. The competent cells were transformed with pIND4 RC1 plasmid coding for RC-LH1 [30]. The transformant strain $RC_x^r$ pIND4 RC1 was grown at 30 °C in RLB medium [30], a low salt Luria-Bertani (LB) medium supplemented with 160 mg of $MgCl_2$, 60 mg of $CaCl_2$ per litre. RLB medium was additionally supplemented with 10 mL of SL-4 micronutrients (cf. DSMZ medium 462) and 2 mL of vitamins (cf. DSMZ medium 462) per litre. The antibiotic kanamycin was added to a final concentration of 25 $mg.L^{-1}$. Batch cultures inoculated into a 2 L volume bioreactor were illuminated by an incandescent tungsten filament light bulb providing white light illumination of 20 μmol photon $m^{-2}$ $s^{-1}$. The culture was aerated by bubbling sterile air (PTFE filter, 0.2 μm pore size) and stirring with a magnetic stirring bar (<100 rpm). Expression of the RC-LH1 encoded by the pIND4 RC1 plasmid was induced by addition of 1 mM IPTG.

***Procapsid isolation and protein purification***

*Procapsid isolation*. *E. coli* cultures were harvested at the mid to late log phase by centrifugation (30 min, 4,500×*g*, 4 °C). The pellet was re-suspended in 50 mM Tris-HCl pH 7.6, 25 mM NaCl, 10 mM $MgCl_2$, with the addition of ethylenediaminetetraacetic acid (EDTA) free complete protease inhibitor cocktail (Merck, Germany), DNAse (10 μg.$mL^{-1}$), RNAse (10 μg.$mL^{-1}$), and thoroughly homogenized. Cells were broken by passage through a cell homogenizer (Microfluidizer, USA) at 172 MPa (25,000 PSI). The unbroken cells and cell wall fragments were separated from the cytosolic fraction containing P22 procapsids by pelleting via low-speed centrifugation (10 min, 5,000×*g*, 4 °C). The supernatant was then pelleted by ultracentrifugation in an Optima XPN-90 ultracentrifuge (Beckman, Japan) using a 70Ti rotor at 45,000 rpm for 60 minutes at 10 °C. The pellet was resuspended in 50 mM Tris-HCl pH 7.6, 25 mM NaCl, 1 mM EDTA, 1mM phenylmethanesulfonyl fluoride (PMSF), loaded on top of a discontinuous sucrose density gradient (top and lower half 0.3 and 0.75 M sucrose respectively in 50 mM Tris-HCl pH 7.6, 25 mM NaCl, 1 mM EDTA, 1mM PMSF) and ultracentrifuged for 1 hour at 45,000 rpm using a 70Ti rotor at 10 °C.

*Coat protein isolation and purification*. Pelleted procapsids were resuspended in 50 mM Tris-HCl pH 7.6, 25 mM NaCl, 1 mM EDTA, 1mM PMSF, 1 mM dithiothreitol (DTT) and 0.8 M guanidine-HCl (GuHCl), loaded on top of a sucrose gradient (0-0.75 M sucrose in 50 mM Tris-HCl pH 7.6, 25 mM NaCl, 1 mM EDTA, 1mM PMSF, 1 mM DTT, 0.8 M GuHCl) and ultracentrifuged for 1 hour at 35,000 rpm using an SW32 rotor at 10 °C. Pelleted, SP-free shells were resuspended in 2 mL of 50 mM Tris-HCl pH 7.6, 25 mM NaCl, 1 mM EDTA, 1mM PMSF, 1 mM DTT, and stored at 4 °C.

*Scaffolding protein isolation and purification*. The supernatant from the pelleted CP procapsid shells contained modified Gp8 scaffolding protein (SP) truncated to amino acid residues 141-303, resulting in an overall mass of 20.3 kDa [23]. The SP was further purified in an Econo FPLC system (Bio-Rad, USA) using 1.5x25 cm column packed with 25 mL of Ni nitrilotriacetic acid (NTA) His Bind Resin (Novagen, USA) equilibrated with binding buffer (50 mM Tris-HCl pH 7.6, 500 mM NaCl, 150 mM imidazole, 1mM PMSF, 1 mM DTT). Samples were loaded onto the column at 2 $mL.min^{-1}$ and washed with the binding buffer until clear outflow, then washed with one column volume of high salt buffer (50 mM Tris-HCl pH 7.6, 1 M NaCl, 150 mM imidazole, 1mM PMSF, 1 mM DTT) to remove nucleic acids bound to the SP. The SP was then eluted with elution buffer (50 mM Tris-HCl pH 7.6, 500 mM NaCl, 500 mM imidazole, 1mM PMSF, 1 mM DTT). Collected fractions were concentrated using a 10 kDa cut off centricon. Final purification was accomplished by size exclusion chromatography (SEC) in an Akta Pure M2 chromatography system (GE Healthcare, Sweden) using a 10x300 mm Superdex75 (Cytiva, USA) SEC column, a single injection of 0.5 mL, and eluted with 0.7 $mL.min^{-1}$ flow by 50 mM Tris-HCl pH 7.6, 250 mM NaCl, 1mM PMSF, 1 mM DTT. The main peak fractions were pooled and concentrated using Amicon devices (Sigma) with a 10 kDa cut-off and stored at 4 °C for immediate use or flash-frozen in 50 μl aliquots in liquid nitrogen and stored at -80 °C.

***RC-LH1 isolation and purification***

*C. sphaeroides* strain $RC_x^r$ pIND4 RC1 cell cultures were harvested at the mid to late log phase by centrifugation (30 min, 4,500×*g*, 4 °C) yielding *ca.* 20 g of wet weight biomass. The pellet was re-suspended in 10 mM Tris-HCl pH 8.0, 150 mM NaCl, 2mM $MgCl_2$, 2mM $CaCl_2$, with addition of DNAse (10 μg.mL$^{-1}$) and RNAse (10 μg.mL$^{-1}$) and thoroughly homogenized. Cells were broken by passage through a cell homogenizer (Microfluidizer, USA) at 172 MPa (25,000 PSI). The unbroken cells and cell wall fragments were separated from the cytosolic fraction and chromatophore vesicles by pelleting via low-speed centrifugation (10 min, 10,000×*g*, 4 °C). The chromatophores recovered from the supernatant were then pelleted by ultracentrifugation (45,000 rpm, 2 h, 4 °C) using a Beckman 70Ti rotor. Pelleted membranes were resuspended in 10 mM Tris-HCl pH 8.0, 150 mM NaCl, 2mM $MgCl_2$, 2mM $CaCl_2$ and diluted to $OD_{590}$ = 10 cm$^{-1}$ corresponding to the Qx band of BChl *a*. The membranes were solubilized by adding DDM to 2 % while stirring in the dark at room temperature for 30 minutes. The insoluble material was then pelleted by ultracentrifugation (45,000 rpm, 30 min, 4 °C) using a 70Ti rotor. Five mL of the solubilized membranes were then layered on top of a continuous sucrose gradient (0.2-2.5M sucrose in 10 mM Tris-HCl pH 8.0, 150 mM NaCl, 2 mM $MgCl_2$, 2 mM $CaCl_2$, 0.025 % DDM) and ultracentrifuged overnight (30,000 rpm, 16 h, 4 °C) using an SW-32Ti rotor. The deep purple band containing crude RC-LH1 was recovered from the ultracentrifuge tube using a Gradient Station ip (Biocomp, Canada). The crude RC-LH1 complexes were further purified using an Econo FPLC system (Bio-Rad, USA) and 1.5x50 cm column packed with 85 mL of Ni-NTA His Bind Resin (Novagen, USA) equilibrated with binding buffer (10 mM Tris-HCl pH 8.0, 150 mM NaCl, 1 mM imidazole, 0.025% DDM). Samples were loaded onto the column at 2 mL.min$^{-1}$ and washed with the binding buffer until clear outflow. The RC-LH1 complexes were then eluted with elution buffer (10 mM Tris pH 8.0, 150 mM NaCl, 150 mM imidazole, 0.025 % DDM). The main elution peak was collected and concentrated using a 100 kDa cut off Amicon device (Sigma). Final purification was accomplished by SEC in an Akta Pure M2 chromatography system (GE Healthcare, Sweden) equipped with a multiple angle laser light scattering (MALS) detector (Wyatt Technologies Heleos II) with dynamic light scattering option (DLS), refractive index (RI) detector (Wyatt Technologies TRex) and a UV-VIS detector. The sample was purified and desalted using a 7.8x300 mm TSKGEL G6000 PWXL (Supelco, Japan) SEC column at a flow rate of 0.5 mL $min^{-1}$ at 23 °C using a mobile phase of 10 mM Tris-HCl pH 8.0, 150mM NaCl, 0.025% DDM. Large scale purification was done using 26x600 mm Superdex 200 (Cytiva, USA) SEC column (320 mL column bed volume), a single 10 mL injection using the superloop, and eluted at 3.5 $mL.min^{-1}$. The main peak fractions were pooled, concentrated using an Amicon device (Sigma) with 100 kDa cut-off to an $OD_{876}$ >100 $cm^{-1}$, and stored in the dark at 4 °C.

***Scaffolding protein coupling to NTA-maleimide linker***

The 6×His tag of the SP was cleaved off by Super Tobacco Etch Virus (TEV) protease (pET29b-10xHis-Super TEV was a gift from EPFL Protein Production and Structure Core Facility (PTPSP) & Florence Pojer (Addgene plasmid # 193833; http://n2t.net/addgene:193833; RRID:Addgene_193833)). The purified SP was mixed with the Super TEV protease in 100:1 molar ratio and allowed to react at room temperature of 3 hours while shaking. Cleaved SP was purified from the undigested protein and Super TEV protease by flowing through HisTrap HP 5 mL Ni-NTA chelating chromatography column (Cytiva USA) equilibrated with binding buffer (50 mM Tris pH 7.6, 250 mM NaCl, 1 mM PMSF, 1 mM DTT). The collected SP was dialyzed against 1 L of 20 mM HEPES, pH 7.6, 250 mM NaCl, 1 mM EDTA. The cleaved SP was then mixed with molar equivalent of NTA-maleimide bifunctional linker (product #12609, AAT Bioquest, Inc., USA) and shaken for 2.5 hours at room temperature and then dialyzed against the HEPES buffer overnight at 4 °C.

***Cytochrome c coupling to NTA-succinimidyl ester linker***

10 mM NTA-succinimidyl (NHS) ester linker (product 12607, AAT Bioquest, Inc., USA) dissolved in dimethylsulfoxide (DMSO) was added gradually in 10 µL steps (134 µL in total) to 500 µL of phosphate buffered saline (PBS) containing 1.3 mM Cyt c from *Equus caballus* heart (Merck, Germany) while shaking for 2.5 hours at room temperature and dialyzed against PBS overnight at 4 °C.

**Procapsid assembly and cargo incorporation**

First, the previously published conditions for procapsid assembly with the SP (fragment 141-303) [27] were adapted for co-assembly with detergent solubilized RC. Monomeric CP was obtained by refolding denatured SP-free shells from 6M guanidinium hydrochloride using overnight dialysis against 50 mM Tris-HCl pH 7.6, 25 mM NaCl, 1 mM EDTA buffer at 4 °C. Aggregates were removed by centrifugation at 40000 RPM in Ti 70 rotor (Beckman Instruments) and the monomeric state was confirmed by HLPC using TSKGEL G4000 PWXL (Supelco, Japan) SEC column at a flow rate of 0.5 mL min$^{-1}$ at 23 °C in mobile phase buffer (50 mM Tris-HCl pH 7.6, 25 mM NaCl, 1 mM EDTA).

For assembly CP (concentration after refolding between 0.6 and 1 mg/ml) and untagged SP were kept at 2:1 molar ratio which is sufficient for assembly but leaves enough volume for SP-linked RC-LH1 and other components. All reactions were performed at 25°C in 50 mM Tris-HCl pH 7.6, 25 mM NaCl buffer supplemented with 0.025% DDM. Assembly reactions (200 μl) were incubated for 2 h after which turbidity did not change. The assembly products were loaded on a TSK-6000 size exclusion column which allowed separation of aggregates and PC like particles from unassembled components. The column was coupled to the multiwavelength absorbance detector of the Akta PURE system and RI MALS/DLS for monitoring the concentrations of components and masses of assemblies.

SP-NTA was mixed with an equimolar amount of $NiCl_2$ and incubated with His-tagged RC-LH1 for 10 min at 25°C. The complex was then mixed with SP and CP to initiate assembly as described above. During optimization the sequence of addition of SP and CP was varied, and the molar ratio of input RC-LH1 to CP was systematically increased from 1:100 (representing ~ 4 RC-LH1 per final capsid) to 1:10. HPLC separation and absorbance at 500 nm were used to monitor RC-LH1 incorporation which was then confirmed by quantitative proteomics of the fractionated samples.

***Electron microscopy***

Assembled products from HPLC fractions were deposited on glow-discharged carbon-coated copper grids and negatively stained with 1.5% uranyl acetate. The grids were visualized in JEOL JEM-2100F transmission electron microscope (JEOL, Japan) at 200 kV and magnification ranging from 35000X to 50000X.

### *Mass spectrometry*

Protein mass was determined using MALDI-TOF/TOF Autoflex mass spectrometer (Bruker Daltonik, USA). Prior the MS experiment, the proteins were purified using C18 ZipTip (Merck, Germany). One-half μL of the intact protein was spotted on the MALDI target and mixed with the same volume of matrix (30 mg.mL$^{-1}$ sinapinic acid dissolved in acetonitrile:0.1% TFA, 7:3). Mass spectra were acquired in the positive linear ion mode using pulsed extraction with an acceleration voltage of 19.5 kV, an extraction voltage 18.2 kV, a lens voltage of 7 kV and delayed extraction time of 360 ns. Resulting spectra were accumulated from up to 500 laser shots. Mass spectra processing was performed using flexAnalysis 3.4 (Bruker Daltonik, USA).

### *Stoichiometry analysis by quantitative proteomics analysis*

Assembled capsids after HPLC separation were processed by in-solution digestion with trypsin, followed by peptide cleanup according to a previously established protocol [43]. Resulting peptides were reconstituted in 40 µl of 3% acetonitrile containing 0.1% formic acid and subjected to nanoLC–MS/MS analysis using an Ultimate 3000 RSLCnano system (Thermo Fisher Scientific) coupled online to a timsTOF Pro mass spectrometer (Bruker Daltonics). Data acquisition was performed in dia-PASEF mode with positive ion polarity, as described previously [44].

Protein identification and quantification were carried out using MaxQuant software (version 2.6.5.0) against a database containing P22 phage and *C. sphaeroides* protein sequences, the *E. coli* K-12 proteome obtained from UniProt (downloaded 25 June 2024), and the contaminant database supplied with MaxQuant [45]. Database searching and data processing parameters were applied as described previously [44].

***Spectroscopy***

*Steady state absorption spectroscopy*. Absorption spectra were measured in a UV-VIS spectrophotometer Lambda 35 (Perkin Elmer, USA) in the range of 250-1000 nm.

*Micro-nanosecond absorption spectroscopy*. Difference absorption spectroscopy was performed using a modified kinetic spectrometer [46]. The light-induced charge separation of the RC-LH1 core complex was driven by broadband pulses (~ mJ energy per pulse) of 2 μs duration delivered by a pair of xenon flash lamps. Absorption spectra were recorded in the 750–980 nm region using a photodiode array detector. The amount of pulse-induced oxidized primary donors ($P_{870}^{+}$) was measured directly by bleaching the main absorption band at 870 nm ($\Delta A_{870} - \Delta A_{955}$). The kinetics of reduction of the oxidized primary donor of RC following the single-turnover saturating pulse were collected by repeated measurements with increasing delay between the actinic and the probe pulses. Alternatively, 3 s pulses from a continuous 850 nm light source (M850L2, Thorlabs, USA) were used to drive charge separation in the RC-LH1 system in some experiments as described in Results. Reduction of the oxidized primary donor of the RC was facilitated by a sodium ascorbate/phenazine methosulfate (PMS) redox system.

ASSOCIATED CONTENT

Supporting Information. Supporting information (PDF) is available and contains Supplementary Figures: (S1) MALDI MS spectra of toolkit components conjugated to bifunctional linkers. (S2) MS spectrum of Cyt c conjugated to NTA-succinimide ester bifunctional linker. (S3) HPLC analyses of assembly components and products. (S4) Negative stain EM galleries of assembly products from HPLC peak fractions. (S5) Proteomics-derived ratios (to coat protein) for proteins in the HPLC purified fractions containing assembled shells. (S6) Additional transient absorption results.

## AUTHOR INFORMATION

### Corresponding Author

*To whom correspondence shall be addressed: Roman Tuma, e-mail: rtuma@prf.jcu.cz, University of South Bohemia, Faculty of Science, Branišovská 1760 370 05, České Budějovice, Czech Republic.

### Author Contributions

D.K., H.L., J.P. and R.T. conceived and designed the experiments, D.K., G.M.W, J.K. and R.T. purified, engineered the proteins and characterized the assemblies, J.T.B provided genetic constructs, microbial strains and expert advice, F.D., J.D. and M.B analyzed the assemblies, D.B. designed, performed and interpreted time resolved experiments, D.K., J.P., D.B and R.T. wrote the manuscript, all authors read and commented on the final text.

### Notes

The authors declare no competing financial interest.

## ACKNOWLEDGMENT

The authors thank to Zdenko Gardián and Eva Ďurinová for TEM imaging. This research was supported by Czech Science Foundation (GACR 22-17333S) (DK, GMW, JG, MB, JD, HL, JP,

RT) , and Canadian NSERC Discovery grant (RGPIN 2018-03898) to JTB. We acknowledge the BC CAS core facility LEM supported by MEYS CR (LM2023050 Czech-BioImaging and OP VVV CZ.02.1.01/0.0/0.0/18_046/0016045).

**Modular molecular toolkit for photochemical energy conversion in a self-assembling nanocontainer**

**Supplementary material - Figures**

David Kaftan[1,2], David Bína[1,3], Guy Michel Wolf[1], Martin Baroch[4,5], Jakub Ködel[1], Juraj Dian[4,5], Filip Dyčka[1], J. Thomas Beatty[6], Heiko Lokstein[5], Jakub Pšenčík[5*] and Roman Tuma[1*]

[1] University of South Bohemia, Department of Chemistry, Branišovská 1760, 37005 České Budějovice, Czech Republic

[2] Czech Academy of Sciences, Institute of Microbiology, Centre Algatech, Novohradská 237, 37981 Třeboň, Czech Republic

[3] Czech Academy of Sciences, Biology Centre, Institute of Plant Molecular Biology, Branišovská 31, 37005 České Budějovice, Czech Republic

[4] Charles University, Faculty of Science, Department of Analytical Chemistry, Albertov 6, 12843 Prague, Czech Republic

[5] Charles University, Faculty of Mathematics and Physics, Department of Chemical Physics and Optics, Ke Karlovu 3, 12116 Prague, Czech Republic

[6] The University of British Columbia, Department of Microbiology & Immunology, Vancouver, V6T 1Z3, Canada

*Corresponding author: Roman Tuma, rtuma@prf.jcu.cz

## SUPPLEMENTARY FIGURES

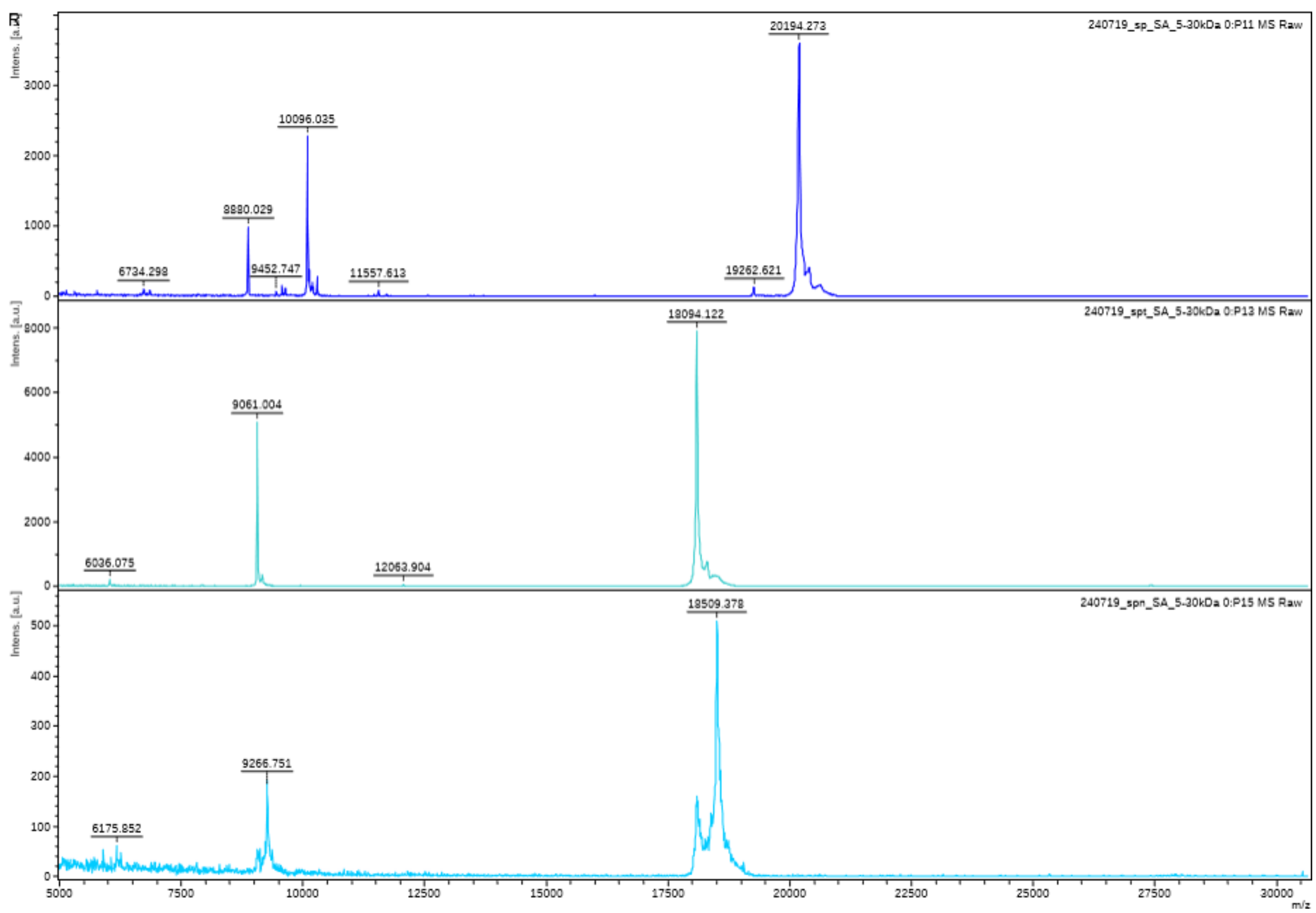

**Supplementary Figure S1.** MALDI MS spectra of toolkit components conjugated to bifunctional linkers. Upper panel: SP in its native form, the measured mass of 20194 Da corresponds to the mass 20185.83 Da expected for a gene product missing the first three residues. Middle panel: SP cleaved by SuperTEV protease, the measured mass 18094 Da corresponds well to the mass expected after cleavage 18110.6 Da. Bottom panel: SP conjugated with NTA-maleimide bifunctional linker. The observed mass of the main peak (singly charged species) 18509 Da corresponds well to the 18524 Da expected for the NTA conjugate. The minor peak shoulder with lower mass indicates presence of approximately 20% of unmodified SP.

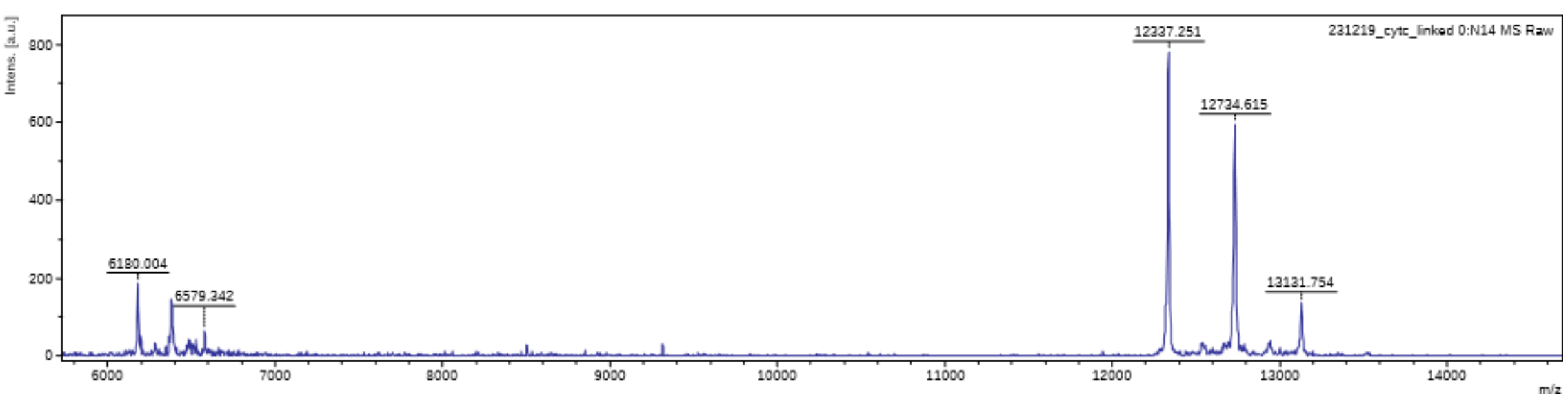

**Supplementary Figure S2.** MS spectrum of Cyt *c* conjugated to NTA-succinimide ester bifunctional linker. The highest intensity peak exhibits an experimental mass of 12337 Da, which is within experimental error of the 12327 Da mass expected for the unmodified bovine Cyt *c*. Higher mass peaks (12734.5 and 13131.8 Da) are observed at mass increments of 397 Da which corresponds to the expected addition of NTA-NHS moieties (399 Da each).

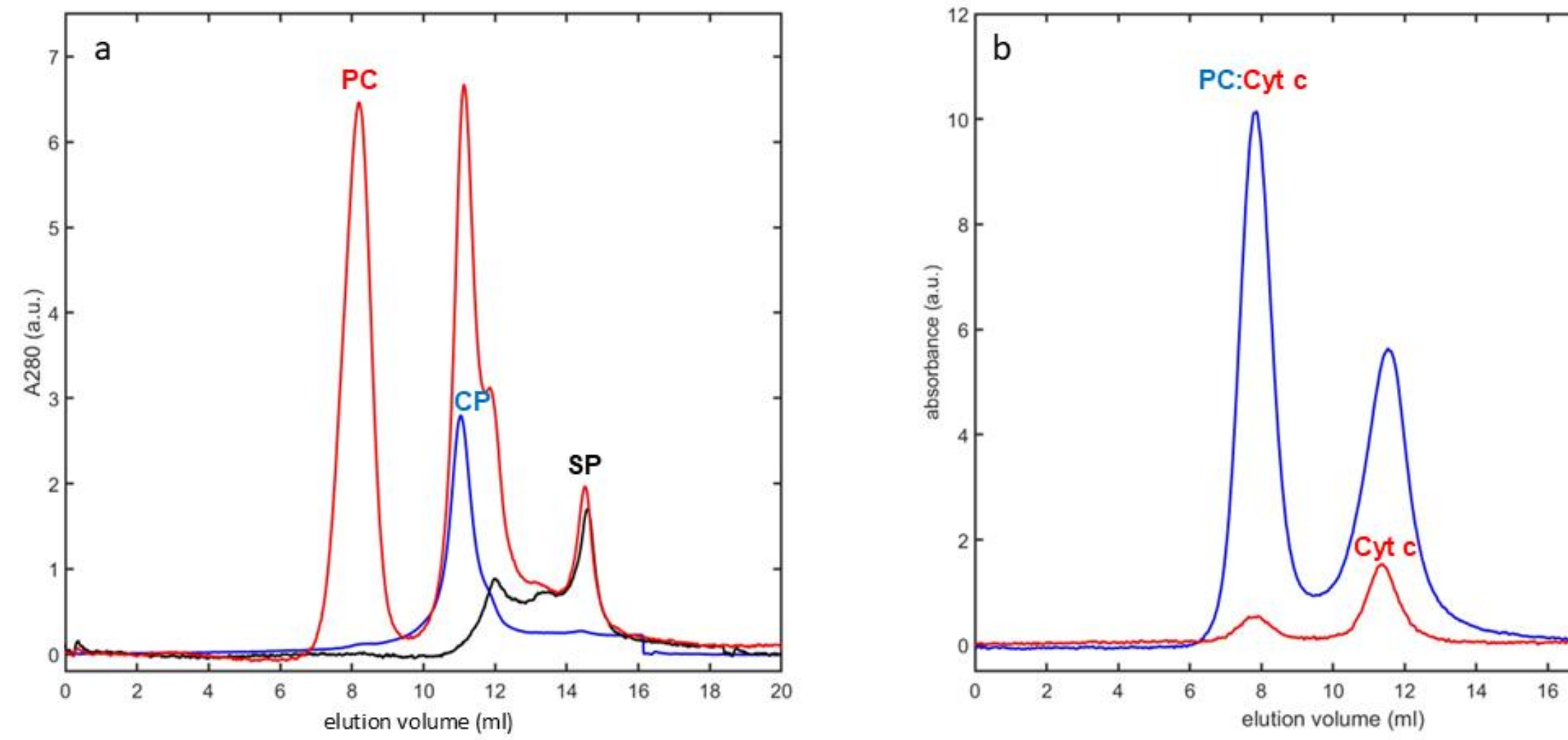

**Supplementary Figure S3.** HPLC analyses of assembly components and products. (a) HPLC analysis (TSK6000 column) of P22 assembly products (red), refolded coat protein (CP, blue) and scaffolding protein (SP, black) monitored by absorbance at 280 nm. Note that SP exhibits two peaks corresponding to monomer (eluting at 14.5 ml) and dimer (12 ml), broadened by a rapid equilibrium (broad peak at ~15 ml). Assembly products contain procapsid shells (PC peak) together with unassembled components at low concentration. (b) HPLC analysis of procapsids assembled in the presence of the Cyt *c* -SP conjugate monitored by absorbance at 280 nm (blue - proteins) and 550 nm (red – heme in Cyt *c*).

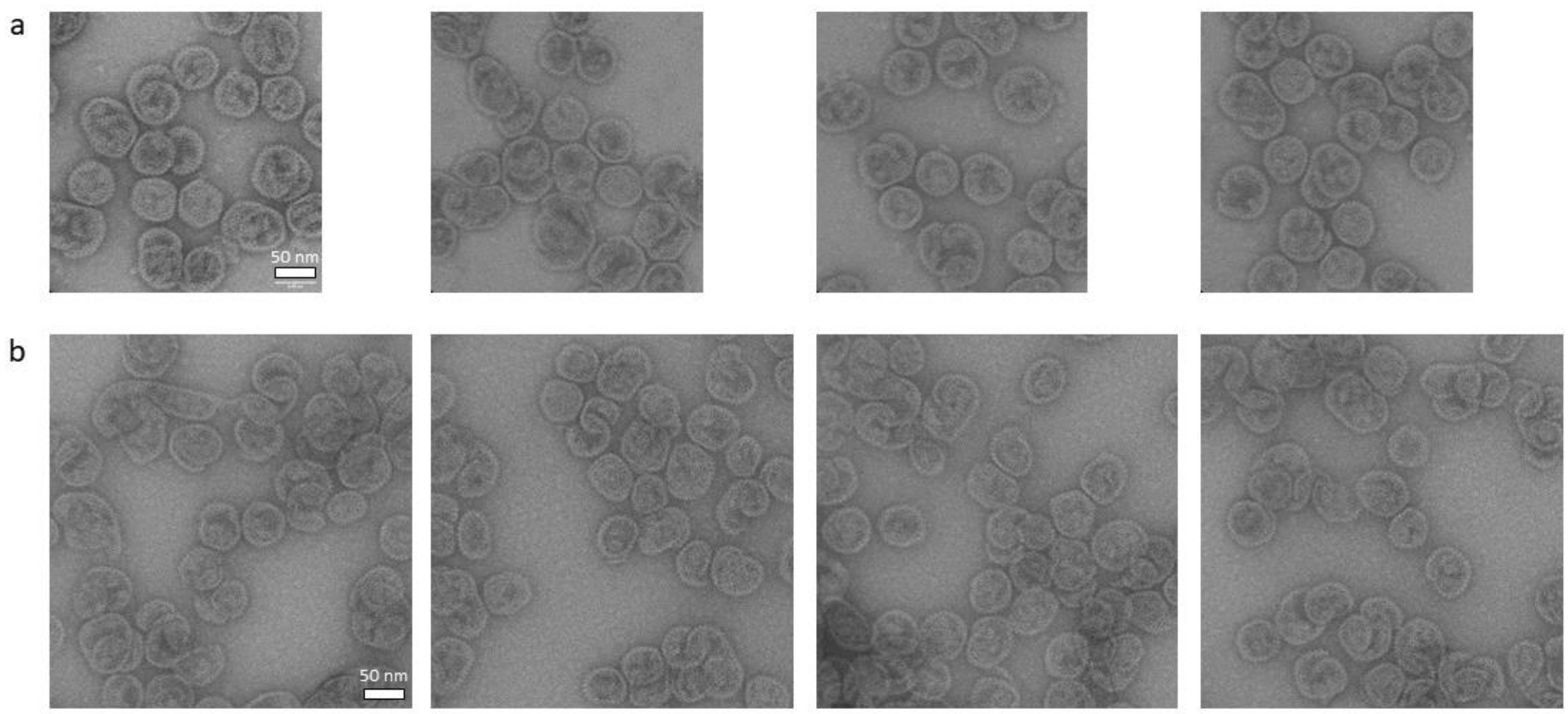

**Supplementary Figure S4.** Negative stain EM galleries of assembly products from HPLC peak fractions in Figure 2. (a) PC:RC-LH1 peak-Fig. 2a. (b) PC:RC-LH1:Cyt c peak-Fig. 2b. Note that micrographs in panels a and b were collected with different magnification hence the fields of view (and figure sizes) differ when shown on the same scale.

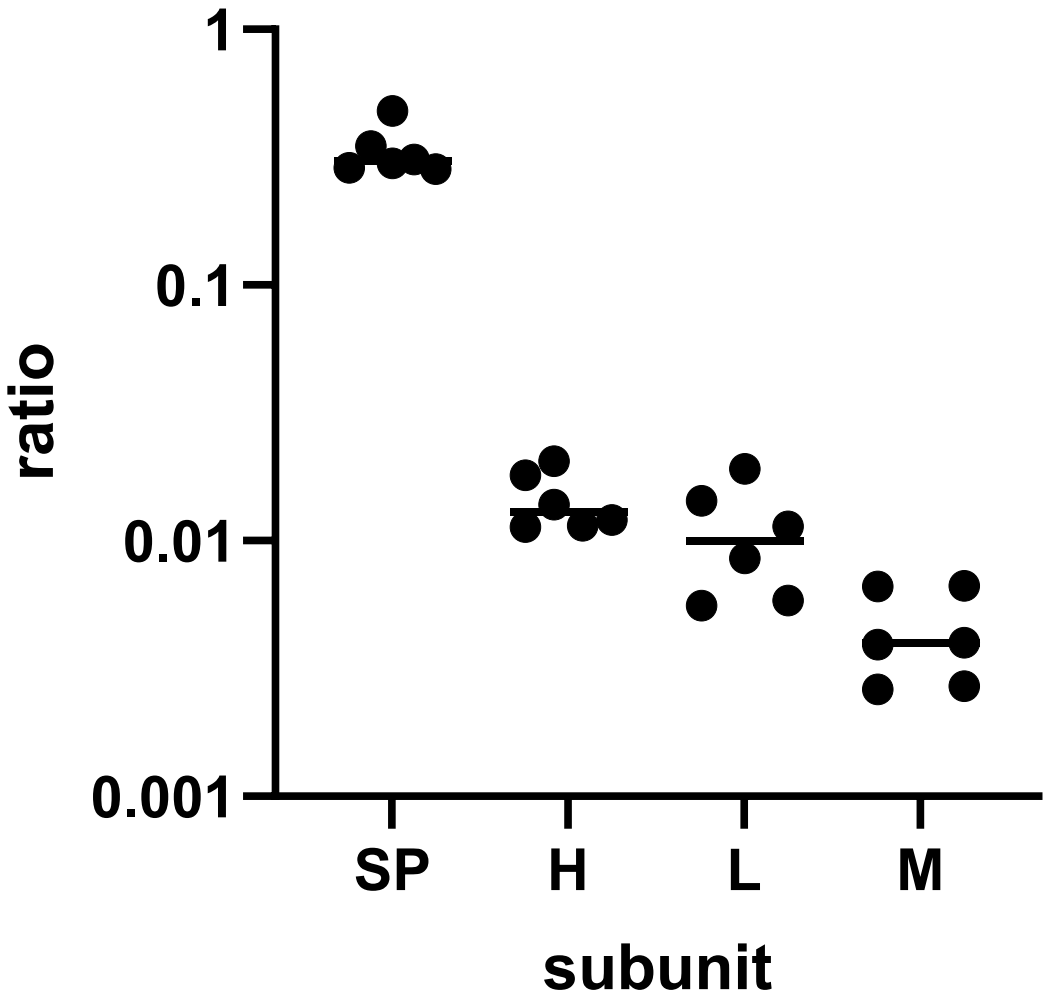

**Supplementary Figure S5.** Proteomics-derived ratios (to coat protein) for proteins in the HPLC purified fractions containing assembled shells. SP – scaffolding to coat ratio, H – RC H subunit to coat, L –RC L subunit to coat, M – RC M subunit to coat.

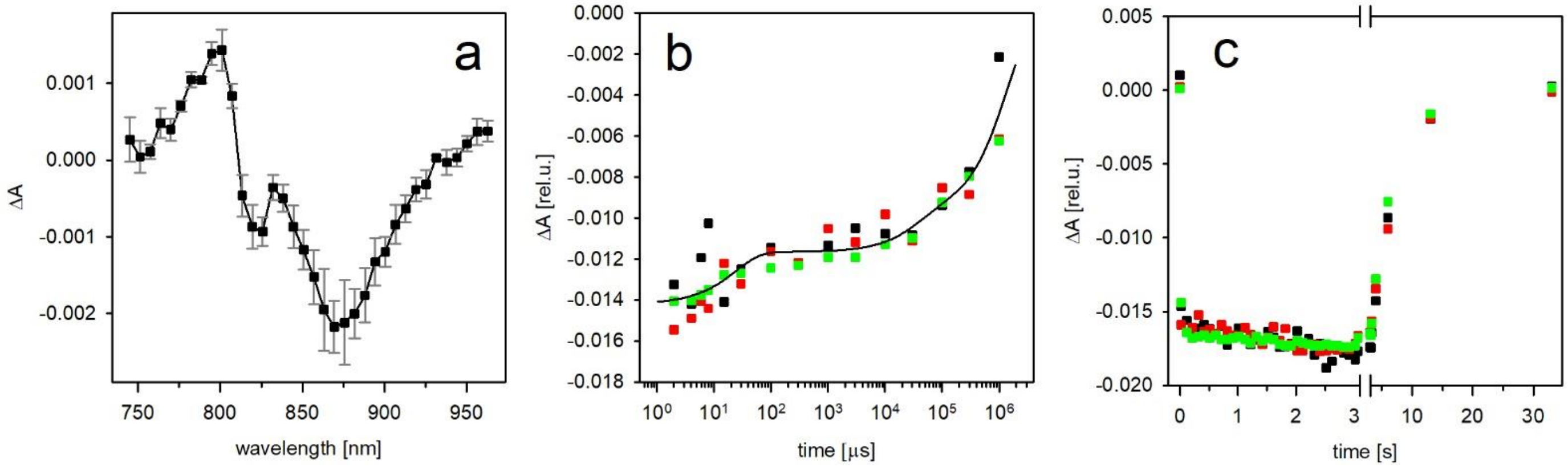

**Supplementary Figure S6.** The transient absorption experiments reported in Fig. 3 of the main text were obtained on samples of different concentrations (hence absorbance, in the range 0.1-0.8). Since the amplitude of the difference absorption signal scales with concentration, the average values and their errors are not useful metrics. Consequently, representative datasets rather than mean±SD of replicated experiments are shown in Figure 3. Here we illustrate the reproducibility of the spectra and kinetics. (a) Flash-induced absorption change of the RC, measured at 3 µs after the actinic flash. Data correspond to a mean ± RMS noise of 4 technical replicates. The noise is well below 0.0002, in agreement with our earlier report [46]; (b and c) show the comparison three biological replicates of measurement of light-induced absorption changes of the RC; (b) kinetics corresponding to the reduction of the light-oxidized primary donor following a 2 ms actinic xenon flash; (c) and time course of absorption change induced by the 3 s LED light pulse. The kinetics were acquired on sample of $OD_{870}$ of ~0.1, ~0.4 and ~0.8 (black, red, green) and rescaled.